\documentclass[aps,prx,onecolumn,superscriptaddress]{revtex4-2}

\usepackage[utf8]{inputenc} % allow utf-8 input
\usepackage[T1]{fontenc}    % use 8-bit T1 fonts
\usepackage{xcolor}         % colors
\usepackage{url}            % simple URL typesetting
\usepackage{booktabs}       % professional-quality tables
\usepackage{amsfonts}       % blackboard math symbols
\usepackage{nicefrac}       % compact symbols for 1/2, etc.
\usepackage{microtype}      % microtypography
\usepackage{braket}
\usepackage{graphicx}
\usepackage{amsmath}
\usepackage{amssymb}
\usepackage{amsthm}
\usepackage[caption=false]{subfig}
\usepackage{algorithmic}

\makeatletter
\let\c@algorithm\relax % Just in case it's defined
\newcounter{algorithm}
\newenvironment{algorithm}[1][htpb]{%
  \begin{figure}[#1]%
  \let\c@figure\c@algorithm
  \refstepcounter{algorithm}%
  \hrule height .8pt \vspace{5pt}%
  \def\caption##1{%
    \noindent \textbf{Algorithm \thealgorithm} ##1 \par
    \vspace{3pt} \hrule height .4pt \vspace{5pt}%
  }%
}{%
  \vspace{5pt} \hrule height .8pt
  \end{figure}%
}

\renewcommand{\p@subsection}{}
\renewcommand{\p@subsubsection}{}

\def\section{%
  \@startsection
    {section}%
    {1}%
    {\z@}%
    {0.8cm \@plus1ex \@minus .2ex}%
    {0.5cm}%
    {\normalfont\large\bfseries\raggedright}%
}
\def\subsection{%
  \@startsection
    {subsection}%
    {2}%
    {\z@}%
    {0.8cm \@plus1ex \@minus .2ex}%
    {0.5cm}%
    {\normalfont\normalsize\bfseries\raggedright}%
}
\def\subsubsection{%
  \@startsection
    {subsubsection}%
    {3}%
    {\z@}%
    {0.8cm \@plus1ex \@minus .2ex}%
    {0.5cm}%
    {\normalfont\normalsize\itshape\raggedright}%
}
\makeatother

\usepackage{graphics}
\usepackage{epsfig}
\usepackage{wrapfig}
\usepackage{bbold}
\usepackage{appendix}
\usepackage{textcomp}
\usepackage{mathrsfs}
\usepackage{dsfont}
\usepackage{bbm}
\usepackage{verbatim}
\usepackage{tikz}
\usepackage{bm}
\usepackage{xr}
\usepackage[
    colorlinks=true,
    citecolor=blue,
    linkcolor=red,
    urlcolor=teal,
    filecolor=magenta
]{hyperref}

\usepackage[most]{tcolorbox}

\newtcolorbox{theorembox}{
  colback=white,        % background color
  colframe=black!70,    % border color
  boxrule=0.5pt,        % border thickness
  arc=2pt,              % small corner rounding
  left=6pt, right=6pt, top=6pt, bottom=6pt % padding
}

\newtcolorbox{fancytheorembox}[1][]{
  enhanced,
  colback=blue!5,        % light background
  colframe=blue!60!black,% border color
  coltitle=black,        % title text color
  fonttitle=\bfseries,   % title font
  colbacktitle=blue!15,  % title background
  boxrule=0.8pt,         % border thickness
  arc=3pt,               % rounded corners
  left=6pt, right=6pt, top=6pt, bottom=6pt, % padding
  attach boxed title to top center={yshift=-2mm},
  title=#1               % optional title
}

\newtheorem{theorem}{Theorem}[section]
\newtheorem{lemma}[theorem]{Lemma}

\newtheorem*{problem-non}{Problem}

\begin{document}

\title{Generative optimal transport via forward-backward HJB matching}

\author{Haiqian Yang}
\altaffiliation{Equal contribution}
% \email{hqyang@mit.edu}
\affiliation{Department of Mechanical Engineering, MIT, Cambridge, MA 02139}

\author{Vishaal Krishnan}
\altaffiliation{Equal contribution}
% \email{vkrishnan@g.harvard.edu}
\affiliation{SEAS, Harvard University, Cambridge, MA 02138}

\author{Sumit Sinha}
\altaffiliation{Equal contribution}
% \email{ssinha@g.harvard.edu}
\affiliation{SEAS, Harvard University, Cambridge, MA 02138}

\author{L. Mahadevan}
\email{lmahadev@g.harvard.edu}
\affiliation{SEAS, Physics and OEB, Harvard University, Cambridge, MA 02138}

\begin{abstract}
% We develop a non-equilibrium thermodynamic formulation of stochastic optimal transport based on a duality between forward- and backward-in-time Hamilton–Jacobi–Bellman (HJB) equations. We consider stochastic processes constrained to connect prescribed endpoint ensembles while minimizing a pathwise cost functional combining spatial penalties and control effort. 
Controlling the evolution of a many-body stochastic system from a disordered reference state to a structured target ensemble, characterized empirically through samples, arises naturally in non-equilibrium statistical mechanics and stochastic control. The natural relaxation of such a system — driven by diffusion — runs from the structured target toward the disordered reference. The natural question is then: what is the minimum-work stochastic process that reverses this relaxation, given a pathwise cost functional combining spatial penalties and control effort? Computing this optimal process requires knowledge of trajectories that already sample the target ensemble — precisely the object one is trying to construct. 
% We resolve this by establishing a duality via time reversal, under which the backward-in-time dynamics is governed by the gradient of a value function satisfying the forward-in-time HJB equation. Via linearization by the Cole–Hopf transformation and its associated Feynman–Kac representation, this duality enables computation of the optimal transport potential from forward-in-time trajectories without explicit backward simulation. 
We resolve this by establishing a time-reversal duality: the value function governing the hard backward dynamics satisfies an equivalent forward-in-time HJB equation, whose solution can be read off directly from the tractable forward relaxation trajectories. Via the Cole–Hopf transformation and its associated Feynman–Kac representation, this forward potential is computed as a path-space free energy averaged over these forward trajectories — the same relaxation paths that are easy to simulate — without any backward simulation or knowledge of the target beyond samples.
The resulting framework provides a physically interpretable description of stochastic transport in terms of path-space free energy, risk-sensitive control, and spatial cost geometry. We illustrate the theory with numerical examples that visualize the learned value function and the induced controlled diffusions, demonstrating how spatial cost fields shape transport geometry analogously to Fermat's Principle in inhomogeneous media. Our results establish a unifying connection between stochastic optimal control, Schrödinger bridge theory, and non-equilibrium statistical mechanics.
\end{abstract}

\maketitle

\section{Introduction}

Many non-equilibrium physical systems exhibit dynamics that are inherently stochastic and path-dependent, and their control remains a long-standing open problem~\citep{davis2024active,takatori2025feedback,peng2016command}. The evolution of these systems is governed by both active forces (or control), and geometric/energetic constraints that confine motion to physically accessible regions of phase space. Achieving optimal control in such systems therefore requires trajectory-level regulation that balances energy expenditure, stochastic fluctuations, and path feasibility, rather than merely prescribing initial and final distributions. Classical frameworks such as stochastic optimal control and dynamic optimal transport provide principled ways to formalize these objectives by minimizing cost functionals over ensembles of trajectories. Bridging these physical control theories with modern parameterized modeling offers a route to learn to control such trajectory-level dynamics from data while retaining physical fidelity.

Recent advances in non-equilibrium thermodynamics construct stochastic processes that transform a reference ensemble into an empirical target ensemble~\citep{sohl2015deep,song2019generative,song2021maximum}. At their core, these frameworks solve a transport problem: how to steer probability mass from source to target under uncertainty. This naturally invites connections to optimal transport and stochastic control, which offer principled formulations for moving distributions with minimal thermodynamic cost via controlled diffusions.
Most existing formulations, however, do not explicitly trace physical optimality back to pure continuous-time trajectory costs. Score-matching algorithms~\citep{sohl2015deep,song2019generative} estimate backward-time drift fields without optimizing a global system action. Stochastic control techniques~\citep{fleming2006controlled} rigorously align thermodynamic transitions, enforcing trajectory-wise efficiency formulas. Flow matching~\citep{lipman2022flow} aligns marginal distributions across time slices, but lacks trajectory-dependent action bounds. Schrödinger bridge frameworks~\citep{wang2021deep,de2021diffusion} cast transformations as entropy-regularized stochastic transport, operating primarily on macroscopic endpoints.

In this work, we propose a transport control framework rooted directly in stochastic optimal control and dynamic optimal transport. We formulate the transition as a time-reversed control problem with fixed endpoint marginals, where the transport drift is given by the gradient of a value function solving a Hamilton--Jacobi--Bellman (HJB) equation. By reversing time and interpreting this HJB equation as a forward optimal control problem, we enable consistent estimation and synthesis from empirical trajectories. Our method parameterizes a scalar potential \( W(t, x) \) whose gradient defines the optimal vector field, offering interpretable, trajectory-optimal transport grounded in physical first principles. Our key contributions are: \text{(1)} We propose a duality theorem (Theorem~\ref{thm:fwd-bwd-hjb}) that connects ensemble transport to a forward stochastic control problem. Instead of blindly estimating score-fields or simulating backward-time SDEs, we parameterize a scalar potential \( W(s, x) \) satisfying a forward Hamilton--Jacobi--Bellman (HJB) equation.  \text{(2)} We introduce a spatial cost function \( \nu(x) \) to further modulate trajectory concentration and regularity, acting as a refractive index shaping the transport geometry, realizing a stochastic manifestation of Fermat’s Principle on path~space.

% \vspace{-5pt}
\paragraph{Related work.}
% \paragraph{Optimal transport theory.} 
Foundational results in optimal transport (OT) theory underpin our framework. Brenier’s theorem~\citep{brenier1991polar} establishes that OT maps under quadratic cost are gradients of convex potentials, motivating generative transport via potential fields. The Kantorovich dual~\citep{villani2008optimal} formulates OT as a saddle-point problem involving scalar potentials, with stochastic extensions linking to optimal control~\citep{mikami2008optimal}. The Benamou--Brenier formulation~\citep{benamou2000computational} casts OT as a dynamic fluid flow minimizing kinetic energy, introducing a velocity field governed by a Hamilton--Jacobi equation. The JKO scheme~\citep{jordan1998variational} interprets diffusion as gradient flow in Wasserstein space, bridging OT and variational evolution. These perspectives inform our formulation, which synthesizes dynamic OT, stochastic control, and potential-based~transport.

% \paragraph{Stochastic optimal control.} 
In continuous-time stochastic control with quadratic costs and dynamics affine in control, the optimal policy is given by the gradient of a value function solving the Hamilton--Jacobi--Bellman (HJB) equation. A key insight due to Kappen~\citep{kappen2005path} showed that, under exponential transformation (Cole--Hopf), the nonlinear HJB reduces to a linear parabolic PDE solvable via the Feynman--Kac formula. This observation enables sample-based estimation of value functions using path integrals over uncontrolled dynamics, bridging control with probabilistic inference. This leads to connections to Girsanov's theorem and Doob's $h$-transform~\citep{haussmann1986time}, revealing a change-of-measure interpretation of optimal control as shifting path distributions. Another line, rooted in information theory~\citep{ET-ET:12,theodorou2010generalized}, recasts control as inference, framing optimal policies as posterior expectations over trajectories. These perspectives converge in Schrödinger bridge formulations~\citep{chen2016relation,leonard2013survey,chen2021likelihood}, where control problems are framed as KL-regularized transport between endpoint distributions. Parallel developments draw from statistical mechanics, viewing the HJB value function as a path-space free energy and connecting stochastic control to thermodynamic identities such as Jarzynski’s equality~\citep{jarzynski1997nonequilibrium,sagawa2010generalized}. Our work builds on this lineage by leveraging the Feynman--Kac representation for training value functions from forward-time samples, while providing a principled generative modeling framework grounded in stochastic control and dynamic transport.

% \paragraph{Deterministic optimal control.} 
Deterministic control theory provides a foundational perspective via the Pontryagin Maximum Principle~\citep{LSP:18} and the adjoint method of Kelley and Bryson~\citep{kelley1962method}, which compute optimal controls by propagating co-state information backward in time. This pathwise integration of the HJB gradient is equivalent to a method of characteristics, and has recently inspired adjoint-matching approaches in generative modeling~\citep{domingo2024adjoint}. These ideas also extend naturally to stochastic control, where adjoint-based supervision corresponds to backward propagation of cost via Feynman–Kac~expectations.

% \paragraph{Thermodynamic transport and control.}
Recent advances in macroscopic state evolution increasingly build on principles from optimal transport (OT), stochastic control, and non-equilibrium mechanics. Stochastic transport models have been rigorously formulated as stochastic control problems governed by Hamilton–Jacobi–Bellman (HJB) equations~\citep{fleming2006controlled, kappen2005path}, explicitly linking phase-space state evolution to macroscopic action costs. Score-based diffusion models~\citep{holderrieth2024hamiltonian} have increasingly drawn on these frameworks~\citep{berner2022optimal, ghimire2023geometry}. Schrödinger bridge methods~\citep{wang2021deep,de2021diffusion,vargas2021solving} frame optimal state transitions as KL-regularized stochastic control. Feynman–Kac-based inference~\citep{huang2021variational, skreta2025feynman, singhal2025general} provides trajectory-based integrability, often incorporating fundamental Girsanov change-of-measure corrections~\citep{haussmann1986time}. Hydrodynamic formulations computationally leverage OT geometry~\citep{lipman2022flow, liu2023generalized} to construct velocity fields for mass transfer mapped against pathwise optimality. Wasserstein gradient flows, JKO-inspired evolutionary schemes~\citep{cheng2024convergence, choi2024scalable}, and variants~\citep{arjovsky2017wasserstein} offer geometric flow mechanics through continuous diffusion along probability manifold spaces. Foundational physics PDE-solvers increasingly investigate high-dimensional HJB solutions via tensor-train approximations and physics-informed integrators~\citep{sommer2024generative, xu2022poisson, liu2023genphys, xu2023pfgm++}. Consequently, mapping non-equilibrium transitions directly as stochastic optimal control unlocks physically principled trajectory boundaries.

% % \vspace{-5pt}
\section{Forward-backward HJB Matching} \label{sec:fwd_rev_hjb_matching}
\begin{figure}[t]
    \centering
    % \vspace{-0.7em}
    \includegraphics[width=0.98\linewidth]{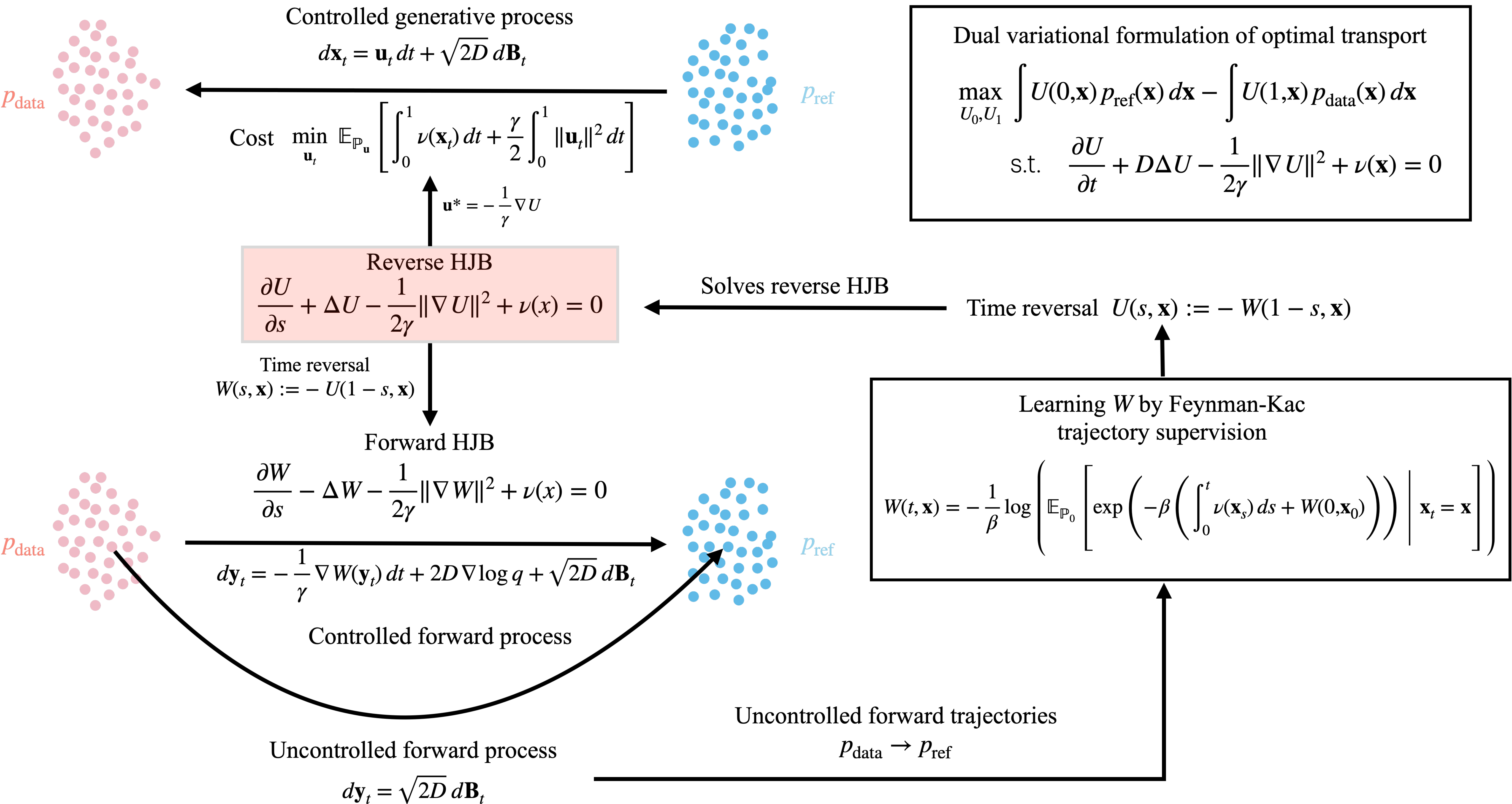}
    \caption{\small
    \textbf{Optimally connecting endpoint marginals via forward-backward HJB matching.} We formulate generative modeling as a stochastic optimal control problem that transports samples from a reference distribution \( p_{\mathrm{ref}} \) to a data distribution \( p_{\mathrm{data}} \), minimizing a trajectory-wise cost. The optimal control policy arises from the backward-time Hamilton--Jacobi--Bellman (HJB) equation for a value function \( U(s, \mathbf{x}) \), whose gradient defines the generative drift. By time-reversing this value function as \( W(s, \mathbf{x}) := -U(1 - s, \mathbf{x}) \), we obtain a forward HJB equation that is solved via a Feynman--Kac path integral, using uncontrolled forward trajectories sampled from \( p_{\mathrm{data}} \to p_{\mathrm{ref}} \). This establishes a dual connection between the forward (training) and backward (generation) dynamics, unified through the value function \( W \) and its governing HJB equation. The formulation incorporates a spatial cost function \( \nu(x) \) that modulates the pathwise transport geometry.}
    \label{fig:schematic}
    % \vspace{-1em}
\end{figure}

% % \vspace{-10pt}
We present a generative modeling framework based on a dual formulation of stochastic optimal control, where the generative drift emerges as the gradient of a value function solving a Hamilton--Jacobi--Bellman (HJB) equation (Fig.~\ref{fig:schematic}). Rather than directly solving the backward-time control problem, which is ill-posed without access to target-to-reference sample trajectories, we construct a forward diffusion process from data to reference, parameterize its value function via a scalar potential, and recover the generative dynamics through time reversal. This forward--backward correspondence yields matched HJB equations connected via the Cole--Hopf transform and Feynman--Kac representation. Our formulation provides a principled mechanism for learning generative flows with controllable trajectory variance and spatial cost geometry, and sidesteps explicit score estimation or backward SDE discretization.

%
% \vspace{-10pt}
\paragraph{Generative modeling based on forward-backward HJB matching.}
We develop a reversible stochastic process framework grounded in stochastic optimal control theory. Our approach constructs both forward and backward stochastic processes as solutions to dual variational problems, avoiding explicit drift estimation and ensuring consistency with optimal transport principles.

% \subsection{Forward Controlled Process}
Let $\mathbf{x}_t \in \mathbb{R}^d$ evolve over the interval $t \in [0,1]$ according to the controlled Itô stochastic differential equation (SDE)
\begin{align} \label{eq:controlled_generative_process}
    d\mathbf{x}_t = \mathbf{u}_t \, dt + \sqrt{2D} \, d\mathbf{B}_t,
\end{align}
where $\mathbf{u}_t$ is the control input, $D > 0$ is the diffusion coefficient, and $\mathbf{B}_t$ is standard Brownian motion. The initial and final distributions are fixed as $p_{\rm ref}$ (reference distribution) and $p_{\rm data}$ (data distribution). We now pose the following control problem
\begin{align} \label{eq:generative_optimal_ctrl_formulation}
    \min_{\mathbf{u}_t} \; \mathbb{E}_{\mathbb{P}_{\mathbf{u}}} \left[ \left. \int_0^1 \nu(\mathbf{x}_t) \, dt + \frac{\gamma}{2} \int_0^1 \| \mathbf{u}_t \|^2 \, dt \; \right| \; \text{s.t.} ~\eqref{eq:controlled_generative_process} \text{~holds~with~} \mathbf{x}_0 \sim p_{\rm ref},~ \mathbf{x}_1 \sim p_{\rm data}  \right],
\end{align}
where $\gamma > 0$ weights the control effort, $\nu(\mathbf{x}) \geq 0$ is a spatial cost function, and $\mathbb{P}_{\mathbf{u}}$ is the path measure induced by the controlled dynamics. The trajectory supervision provided by $\nu(\mathbf{x})$ could be useful in physics and engineering systems, where physically invalid states and safety constraints can be penalized by the spatial cost geometry.

Problem~\eqref{eq:generative_optimal_ctrl_formulation} defines a stochastic optimal control problem with fixed terminal marginals and a running cost comprising both spatial and control penalties. While this problem is posed in terms of controlled path measures, it admits a dual reformulation in terms of a scalar potential solving a nonlinear Hamilton--Jacobi--Bellman (HJB) equation. This dual form not only characterizes the optimal control explicitly, but also underlies our modeling framework by enabling value function learning via forward-time sampling. We state this result in the lemma below (proof in Appendix~\ref{app:proof:generative_opt_ctrl}).

% % \vspace{-10pt}
\begin{lemma}[\bf \emph{Dual variational principle}] \label{lemma:generative_opt_ctrl}
Let \( \gamma > 0 \), \( D > 0 \), let \( p_{\mathrm{ref}}, p_{\mathrm{data}} : \mathbb{R}^d \to \mathbb{R}_{\geq 0} \) be probability density functions, each belonging to \( L^1(\mathbb{R}^d) \cap L^2(\mathbb{R}^d) \) and satisfying \( \int \|x\|^2 p(x) \, dx < \infty \) and let \( \nu : \mathbb{R}^d \to \mathbb{R}_{\geq 0} \) be a measurable, locally bounded (with at most polynomial growth) function. Then the optimal control \( \mathbf{u}^* \) for the SDE~\eqref{eq:controlled_generative_process}, that minimizes the expected total cost in~\eqref{eq:generative_optimal_ctrl_formulation}, is given in feedback form by $\mathbf{u}^*(t, \mathbf{x}) = -\nicefrac{1}{\gamma} \nabla U(t, \mathbf{x})$, where \( U : [0,1] \times \mathbb{R}^d \to \mathbb{R} \) is a solution to the dual variational problem
\begin{align} \label{eq:generative_potential}
    \begin{aligned}
        \max_{U_0, U_1} \; &\int U(0, \mathbf{x}) \, p_{\rm ref}(\mathbf{x}) \, d\mathbf{x} - \int U(1, \mathbf{x}) \, p_{\rm data}(\mathbf{x}) \, d\mathbf{x} \\
        &\text{s.t.} \quad \frac{\partial U}{\partial t} + D \Delta U - \frac{1}{2\gamma} \| \nabla U \|^2 + \nu(\mathbf{x}) = 0.
    \end{aligned}
\end{align}
The controlled SDE~\eqref{eq:controlled_generative_process} with $\mathbf{u}^*(t, \mathbf{x}) = -\nicefrac{1}{\gamma} \nabla U(t, \mathbf{x})$ then yields an optimal transport process (in the sense of~\eqref{eq:generative_optimal_ctrl_formulation}) from $\mathbf{x}_0 \sim p_{\rm ref}$ to $\mathbf{x}_1 \sim p_{\rm data}$ over $[0,1]$. 
\end{lemma}

While Lemma~\eqref{eq:generative_potential} characterizes the optimal control as the gradient of a potential~$U$, its implementation poses a fundamental challenge. The variational objective requires evaluating expectations with respect to the data distribution \( p_{\rm data} \), but doing so via a sampling procedure starting from the reference initialization \( p_{\rm ref} \) presupposes knowledge of the generative process itself. In other words, generating samples from \( p_{\rm data} \) is a prerequisite for learning the very control that enables generation. This circular dependency makes the direct generative formulation intractable. To resolve this, we instead construct a forward diffusion process that flows from \( p_{\rm data} \) to \( p_{\rm ref} \), enabling supervised training of a potential function on accessible samples. The generative process is then recovered via time reversal, using the learned potential to define a drift aligned with the optimal control.

We formalize this time-reversal connection in the following theorem, which establishes the correspondence between the above generative (backward) process and a forward diffusion process, both governed by matched Hamilton-Jacobi-Bellman (HJB) equations. This result underlies our modeling approach, enabling sample generation from a reference distribution via a learned dual potential (see Appendix~\ref{app:fwd-bwd-hjb} for proof).

\begin{fancytheorembox}
\begin{theorem}[\bf \emph{Forward–Backward HJB Duality}] \label{thm:fwd-bwd-hjb}
Let \( W : [0,1] \times \mathbb{R}^d \to \mathbb{R} \) be a forward potential defined by the time reversal of the generative potential~$U$ in~\eqref{eq:generative_potential} as $W(s, \mathbf{x}) := -U(1 - s, \mathbf{x})$. Then:

\noindent \text{(1)} The function \( W \) satisfies the forward HJB equation
\begin{align} \label{eq:forward_hjb}
    \frac{\partial W}{\partial s} - D \Delta W - \frac{1}{2\gamma} \| \nabla W \|^2 + \nu(x) = 0.
\end{align}
    
\noindent \text{(2)} The controlled SDE $d\mathbf{y}_s = \mathbf{v}^*(s, \mathbf{y}_s)\,ds + \sqrt{2D}\,d\mathbf{B}_s$, with the control vector field \( \mathbf{v}^*(s, \mathbf{x}) := -\frac{1}{\gamma} \nabla W(s, \mathbf{x}) + 2D \nabla \log q(s, \mathbf{x}) \) and initial condition \( \mathbf{y}_0 \sim p_{\mathrm{data}} \) transports \( p_{\mathrm{data}} \to p_{\mathrm{ref}} \) over the interval \( s \in [0,1] \).

\noindent \text{(3)} The control field \( \mathbf{v}^* \) solves the forward stochastic optimal control problem
\[
\min_{\mathbf{v}} \; \mathbb{E}_{\mathbb{Q}_{\mathbf{v}}} \left[ \int_0^1 \nu(\mathbf{y}_s)\,ds + \frac{\gamma}{2} \int_0^1 \|\mathbf{v}_s - \mathbf{s} \|^2\,ds \right],
\]
subject to the SDE $d\mathbf{y}_s = \mathbf{v}_s\,ds + \sqrt{2D}\,d\mathbf{B}_s$, with $\mathbf{y}_0 \sim p_{\mathrm{data}}$,  $\mathbf{y}_1 \sim p_{\mathrm{ref}}$, and \( q(s, \mathbf{x}) \) denoting the time-dependent marginal density of \( \mathbf{y}_s \), and \( \mathbf{s}(s, \mathbf{x}) := 2D \nabla \log q(s, \mathbf{x}) \) is the corresponding score correction term arising from the Fokker--Planck dynamics of the controlled process.
\end{theorem}
\end{fancytheorembox}

Some comments on Theorem~\ref{thm:fwd-bwd-hjb} are in order. We note that the generative process \( (\mathbf{x}_t) \) and the forward process \( (\mathbf{y}_s) \) are related through the time-reversal transformation \( W(s, \mathbf{x}) = -U(1 - s, \mathbf{x}) \), and induce identical marginal density evolutions when time is reversed. The dual variational formulation in Lemma~\ref{lemma:generative_opt_ctrl} thereby enables, via time reversal, the learning of a single scalar potential \( W \) from forward-time diffusion trajectories, whose gradient prescribes the optimal control for generative transport in reverse. Crucially, it eliminates the need for explicit score estimation or backward SDE construction as both are implicitly accounted for through the forward stochastic control problem and its associated HJB equation. This yields a variationally consistent mechanism for learning generative~flows.

The structure of the optimal forward control field $\mathbf{v}^*(s, \mathbf{x}) = -\frac{1}{\gamma} \nabla W + 2D \nabla \log q$ in Theorem~\ref{thm:fwd-bwd-hjb} admits a decomposition into two components with distinct physical roles. The first term, $-\frac{1}{\gamma} \nabla W$, drives transport along the gradient of the value function, steering trajectories toward regions of lower accumulated cost-to-go. The second term, $2D \nabla \log q$, is the score of the time-marginal density $q(s, \mathbf{x})$, which provides a diffusive correction ensuring consistency with the Fokker--Planck evolution of the forward process. Together, they decompose the optimal policy into a goal-directed component (value function gradient) and a density-aware correction (score), without requiring either to be estimated independently --- both are encoded in the single scalar potential $W$ through its governing HJB equation and the associated Fokker--Planck dynamics.
%

% \vspace{-5pt}
\paragraph{Feynman--Kac estimation of the forward potential.}
To evaluate the forward potential \( W \) solving the HJB equation~\eqref{eq:forward_hjb}, we apply the Cole--Hopf transformation \( W = \nicefrac{1}{\beta} \log Z \), where \( \beta = \nicefrac{1}{2D\gamma} \). This change of variables linearizes the nonlinear HJB into the parabolic PDE
\begin{align} \label{eq:cole-hopf_linearized}
    \frac{\partial Z}{\partial t} = D \Delta Z - \beta \nu Z,
\end{align}
which has the structure of a diffusion equation with a spatially varying absorption rate $\beta\nu(\mathbf{x})$. Physically, the exponential weighting implied by this absorption term suppresses contributions from paths that traverse high-cost regions of state space, while amplifying those that remain in low-cost corridors. The solution admits a pathwise representation via the Feynman--Kac formula. Specifically, if \( \mathbf{x}_s \) follows standard Brownian motion under the measure \( \mathbb{P}_0 \), i.e., $d\mathbf{x}_s = \sqrt{2D} \, d\mathbf{B}_s$, then the solution to the linear PDE~\eqref{eq:cole-hopf_linearized} is given by
\begin{align} \label{eq:FK_estimate}
    Z(t, \mathbf{x}) = \mathbb{E}_{\mathbb{P}_0} \left[ Z(0, \mathbf{x}_0)  \exp\left( - \beta \int_0^t \nu(\mathbf{x}_s) \, ds \right) \,\middle|\, \mathbf{x}_t = \mathbf{x} \right].
\end{align}

In practice, however, sampling trajectories from pure Brownian motion is inefficient when the data distribution \( p_{\mathrm{data}} \) is far from the reference \( p_{\mathrm{ref}} \). Instead, we simulate trajectories using a reference process with drift, typically a Langevin dynamics
\begin{align} \label{eq:langevin}
    d\mathbf{x}_s = -\nabla V(\mathbf{x}_s) \, ds + \sqrt{2D} \, d\mathbf{B}_s, \quad \mathbf{x}_0 \sim p_{\mathrm{data}},
\end{align}
which induces a different path measure \( \mathbb{P}_V \). While this correction can be derived via Girsanov's theorem,  we instead directly use the original Feynman--Kac expression~\eqref{eq:FK_estimate} and evaluate it over forward trajectories sampled from the Langevin dynamics~\eqref{eq:langevin}. In this case, the mismatch between the sampling and reference measures must be accounted for at generation time by explicitly reversing the Langevin drift. This yields the generative~process
\begin{align} \label{eq:generative_langevin_correction}
    d\mathbf{x}_t = \left( \nabla V(\mathbf{x}_t) - \frac{1}{\gamma} \nabla U(t, \mathbf{x}_t) \right) \, dt + \sqrt{2D} \, d\mathbf{B}_t,
\end{align}
which corrects for the forward drift and generates samples consistent with the original Feynman--Kac formulation.
Taken together, this formulation forms the basis for training the forward potential \( W \), and allows sample-based estimation of the value function using diffusion paths generated from a tractable reference process. In practice, we use this estimator to supervise the output of a neural network approximation of \( Z \), enabling efficient training of the generative potential \( W \) via its Cole--Hopf representation.
Furthermore, a Taylor expansion of the value function \( W \), via the Cole--Hopf transformation \( W = \nicefrac{1}{\beta} \log Z \), reveals that it captures both the expected cumulative cost and its pathwise variance. The inverse temperature \( \beta = 1/(2D\gamma) \) modulates this tradeoff, with smaller \( \gamma \) favoring low-variance trajectories. Thus, variance control emerges naturally from the stochastic optimal control formulation, without requiring explicit regularization\footnote{Refer to Appendix~\ref{app:forward-potential-clarifications} for details on Girsanov correction and risk-sensitive control.}.

% \vspace{-5pt}
\section{Learning generative scalar potential}
% \vspace{-5pt}
% \paragraph{Overview.}
Training our model involves learning a scalar potential \( W_\theta(s, \mathbf{x}) \) that solves a forward Hamilton--Jacobi--Bellman (HJB) equation and determines the generative transport field. To achieve this, we simulate a forward diffusion process from the data distribution to the reference, and use the Feynman--Kac representation to supervise the value function using sample paths. Furhtermore, we jointly optimize a boundary value objective derived from the dual formulation of stochastic control, which involves only evaluations of the potential at the endpoints of the forward process. Taken together, this eliminates the need for score estimation or backward-time SDE simulation. We describe below the construction of training trajectories and the estimation of learning targets.

% \paragraph{Ornstein-Uhlenbeck process.} 
For the forward diffusion, we employ trajectories generated by an Ornstein--Uhlenbeck (OU) process (arising from a Langevin dynamics~\eqref{eq:langevin} with $V(x) = \nicefrac{\theta}{2} \left\| x \right\|^2$) that transports samples from the data distribution \( p_{\rm data} \) to the reference \( p_{\rm ref} \). Specifically, we use the discretized dynamics
\begin{align} \label{eq:OU_process}
    \mathbf{x}_{k+1} = (1 - \theta \Delta s)\, \mathbf{x}_k + \sqrt{2D \Delta s}\, \xi_k, \quad \xi_k \sim \mathcal{N}(0, I),
\end{align}
which corresponds to the OU process \( d\mathbf{x}_s = -\theta \mathbf{x}_s\, ds + \sqrt{2D}\, d\mathbf{B}_s \), with \( \mathbf{x}_0 \sim p_{\rm data} \). These forward-time trajectories are used to supervise learning of the value function via the Feynman--Kac representation. To supervise the potential \( W_\theta \), we typically roll out entire forward paths from \( \mathbf{x}_0 \sim p_{\rm data} \); however, whenever only isolated time slices are needed, we exploit the analytic transition kernel of the OU process $\mathbf{x}_t \mid \mathbf{x}_s \sim \mathcal{N}\left( e^{-\theta (t - s)} \mathbf{x}_s, \; \frac{D}{\theta} \left(1 - e^{-2\theta (t - s)}\right) \mathbf{I} \right)$,
allowing for efficient conditional sampling without simulating full trajectories.

\paragraph{Learning the HJB value function via trajectory supervision.} We train a neural network to represent the HJB value function \( W_\theta(x, s) \), where \( Z = \exp ( \beta W_\theta ) \) satisfies the linear PDE~\eqref{eq:cole-hopf_linearized}. This potential is supervised using forward-time trajectories generated by the OU process, and targets are provided via the Feynman--Kac estimate of the PDE solution, given in~\eqref{eq:FK_estimate}.
% and corrected under Girsanov’s change of measure in~\eqref{eq:FK_estimate_girsanov_corrected}.
We use the forward paths of the OU process and estimate \( Z(s, \mathbf{x}) \) at intermediate points by evaluating the terminal potential and integrating the accumulated cost-to-go. For a trajectory~$i$, consisting of points \( \left \lbrace \mathbf{x}_k^{(i)} \right \rbrace_{k=0}^K \) with uniform time steps \( s_k = k \Delta s \), we let
\begin{align}
    Z^{(i)}_{\mathrm{FK}}(s_k, \mathbf{x}_k^{(i)}) := \exp\left( - \beta \sum_{j=0}^{k-1} \nu_\phi(\mathbf{x}_j^{(i)}) \, \Delta s \right) \exp \left( \beta W_\theta (0, \mathbf{x}_0^{(i)}) \right).
\end{align}
where $Z^{(i)}_{\mathrm{FK}}(s_k, \mathbf{x}_k^{(i)})$ denotes the per-trajectory Feynman--Kac contribution from the $i$-th forward sample path. The full Feynman–Kac estimate of $Z_\theta$ would be obtained by averaging the contributions across the dataset. This is achieved by the corresponding regression loss
\begin{align} \label{eq:FK_loss}
    \mathcal{L}_{\mathrm{FK}} := \frac{1}{NK} \sum_{i,k} \left( \exp \left( \beta W_\theta (s_k, \mathbf{x}_k^{(i)}) \right) - Z^{(i)}_{\mathrm{FK}}(s_k, \mathbf{x}_k^{(i)}) \right)^2.
\end{align}

To encourage short-term consistency, we introduce a local one-step loss between adjacent points
\begin{align} \label{eq:FK_local_loss}
\small
\begin{aligned}
    \mathcal{L}_{\rm FK-local} := \frac{1}{NK} \sum_{i,k} \left(  
     \exp \left( \beta W_\theta (s_{k+1}, \mathbf{x}_{k+1}^{(i)}) \right)
     -  e^{-\beta \nu_\phi(x_k^{(i)}) \Delta s} \exp \left( \beta W_\theta (s_k, \mathbf{x}_k^{(i)}) \right) \right)^2.
\end{aligned}
\end{align}

For the dual variational objective, we consider the following loss which supervises the endpoint values of the potential
\begin{align} \label{eq:dual_loss}
    \mathcal{L}_{\rm dual} := \frac{1}{N} \sum_{i=1}^N W_\theta \left( \mathbf{x}_K^{(i)} \right) - \frac{1}{N} \sum_{i=1}^N W_\theta \left( \mathbf{x}_0^{(i)} \right).
\end{align}
The total training loss is given by the sum of the above losses 
\begin{align} \label{eq:total_loss}
    \mathcal{L}_{\text{total}}(\theta) := 
    \lambda_{\rm FK} \mathcal{L}_{\mathrm{FK}} + \lambda_{\rm FK-local} \mathcal{L}_{\mathrm{FK-local}} + \lambda_{\rm dual} \mathcal{L}_{\mathrm{dual}}.
\end{align}
Each component of the total loss addresses a distinct aspect of the HJB solution structure. The Feynman--Kac loss $\mathcal{L}_{\rm FK}$ enforces global consistency of the parameterized value function with the integrated cost-to-go along entire forward trajectories, ensuring that the learned potential faithfully represents the path-integral solution of the linearized PDE. The local loss $\mathcal{L}_{\rm FK-local}$ complements this by imposing one-step temporal consistency between adjacent time points, acting as a discrete semigroup constraint that stabilizes training and improves gradient signal at fine temporal scales. The dual loss $\mathcal{L}_{\rm dual}$ encodes the boundary conditions of the Kantorovich dual problem (Lemma~\ref{lemma:generative_opt_ctrl}): it encourages the potential to attain high values near the data distribution (where forward-time transport originates at $s = 0$) and low values near the reference (where it terminates at $s = 1$), thereby enforcing the correct directionality of probability flow. When the spatial cost $\nu$ is not prescribed but learned, it is parameterized by a separate network $\nu_\phi(\mathbf{x})$ and optimized jointly with $W_\theta$ through the same loss.
A complete summary of the potential learning and loss computation steps is provided in Algorithm~\ref{alg:HJB_training}.

\begin{algorithm}[htpb]
\caption{Learning the HJB value function via trajectory supervision}
    \begin{algorithmic}[1] \label{alg:HJB_training}
    \REQUIRE Dataset~$\mathcal{D}$, learning rate $\eta$, initial parameters $\theta_0$, loss weights $\lambda_{\rm FK}, \lambda_{\rm FK-local}, \lambda_{\rm dual}$, step size \( \Delta s \), number of steps \( K \)
    \STATE Initialize $\theta \leftarrow \theta_0$ 
    \WHILE{$\theta$ has not converged} 
    \STATE Sample $\mathbf{x}_0$ uniformly at random from $\mathcal{D}$
    \FOR{$t = 1$ to $T$}
        \STATE Evaluate $W_\theta(0, \mathbf{x}_0)$
        \STATE Sample forward trajectory $\left \{ (s_k, \mathbf{x}_k) \right \}_{k=1}^N$ via the OU process~\eqref{eq:OU_process}
        \STATE Evaluate $W_\theta(s_k, \mathbf{x}_k)$ on sample trajectory $\left \{ (s_k, \mathbf{x}_k) \right \}_{k=1}^N$
        \STATE Compute losses $\mathcal{L}_{\mathrm{FK}}$ in \eqref{eq:FK_loss}, $\mathcal{L}_{\mathrm{FK-local}}$ in \eqref{eq:FK_local_loss}, $\mathcal{L}_{\mathrm{dual}}$ in \eqref{eq:dual_loss}
        \STATE Compute total loss $\mathcal{L}_{\mathrm{total}} = \lambda_{\rm FK} \mathcal{L}_{\mathrm{FK}} + \lambda_{\rm FK-local} \mathcal{L}_{\mathrm{FK-local}} + \lambda_{\rm dual} \mathcal{L}_{\mathrm{dual}}$ as in \eqref{eq:total_loss}
        \STATE Update parameters: $\theta \leftarrow \theta - \eta \nabla_\theta \mathcal{L}_{\mathrm{total}}$
    \ENDFOR

    \ENDWHILE
    \STATE \textbf{Return} Optimized parameters~$\theta^*$
    \end{algorithmic}
\end{algorithm}

\paragraph{Sample generation by backward-time controlled diffusion.}
Once the HJB value function \( W_\theta(s, \mathbf{x}) \) has been trained, sample generation proceeds by starting from the reference distribution \( p_{\rm ref} \) and simulating the controlled process $d\mathbf{x}_s = \theta \mathbf{x}_s  + \frac{1}{\gamma} \nabla W_\theta(1-s, \mathbf{x}_s) \, ds + \sqrt{2D} \, d\mathbf{B}_s,$ with $\mathbf{x}_0 \sim p_{\rm ref}$,
where the first drift term corresponds to the reversal of the OU drift, while the second drift term is defined by the gradient of the learned value function \( \nabla W_\theta \). In practice, we sample from this process using the Euler--Maruyama scheme
\begin{align} \label{eq:generative_discrete_implementation}
    \mathbf{x}_{k+1} = \mathbf{x}_k + \Delta s \left( \theta \mathbf{x}_k + \frac{1}{\gamma} \nabla W_\theta(1-s_k, \mathbf{x}_k) \right) + \sqrt{2D \Delta s} \, \xi_k, \quad \xi_k \sim \mathcal{N}(0, I),
\end{align}
where time is discretized as \( s_k = k \Delta s \) with \( s_0 = 0 \), \( s_K = 1 \), and \( \mathbf{x}_0 \sim p_{\rm ref} \). This generative process yields samples approximating the data distribution \( p_{\rm data} \), while following paths optimized for the transport cost structure learned during training. We summarize the generative sampling procedure, based on the time-reversed controlled diffusion~\eqref{eq:generative_discrete_implementation}, in Algorithm~\ref{alg:HJB_generation}.

\begin{algorithm}[htpb]
\caption{Sample generation via backward-time controlled diffusion}
\label{alg:HJB_generation}
\begin{algorithmic}[1]
\REQUIRE Trained potential \( W_\theta(s, x) \), reference distribution \( p_{\rm ref} \), step size \( \Delta s \), number of steps \( K \), parameters \( \theta, \gamma, D \)
\STATE Sample \( \mathbf{x}_0 \sim p_{\rm ref} \)
\FOR{$k = 0$ to $K-1$}
    \STATE Set \( s_k = k \Delta s \)
    \STATE Sample \( \xi_k \sim \mathcal{N}(0, I) \)
    \STATE Update \( \mathbf{x}_{k+1} \) via Process~\eqref{eq:generative_discrete_implementation}
\ENDFOR
\STATE \textbf{Return} \( \mathbf{x}_K \) as generated sample drawn from \( p_{\rm data} \)
\end{algorithmic}
\end{algorithm}

% % \vspace{-10pt}
\section{Results}
% % \vspace{-10pt}
% 
\begin{figure}[htpb]
    \centering
    \includegraphics[width=1.0\linewidth]{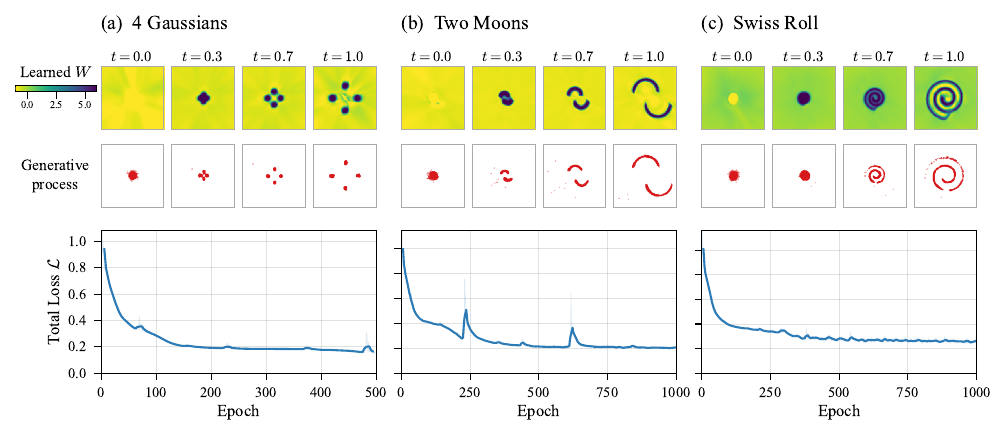}
    \caption{\small
    \textbf{HJB value function learning and generative transport on standard 2D benchmarks.}
    We solve the stochastic optimal control problem (Eqn.~\ref{eq:generative_optimal_ctrl_formulation}): find the control \( \mathbf{u}^*(t,\mathbf{x}) = \nabla_{\mathbf{x}} W \) that transports the reference \( p_{\rm ref} \sim \mathcal{N}(0, I) \) to the empirical target \( p_{\rm data} \) while minimizing a pathwise cost combining spatial penalty \( \nu(\mathbf{x}) \) and quadratic control effort.
    The scalar value function \( W(t, \mathbf{x}) \), which solves the forward HJB equation (Eqn.~\ref{eq:forward_hjb}), is parameterized as a 10-layer MLP (64 hidden units, GeLU activations, sinusoidal positional time embeddings) and trained via Feynman--Kac trajectory supervision using the total loss \( \mathcal{L}_{\rm total} = \mathcal{L}_{\rm FK} + \mathcal{L}_{\rm dual} \) (Eqn.~\ref{eq:total_loss}, Algorithm~\ref{alg:HJB_training}).
    Forward trajectories are generated by an Ornstein--Uhlenbeck process with diffusion coefficient \( D = 0.05 \) and risk parameter \( \beta = 0.1 \), discretized over \( T = 128 \) steps.
    The model is trained for \( n = 2000 \) epochs on \( N = 4096 \) samples using the Adam optimizer (learning rate \( \eta = 10^{-3} \)), implemented in JAX.
    Results are shown on three standard 2D benchmarks used in the generative transport literature~\citep{de2021diffusion,lipman2022flow}:
    \textbf{(a)} 4 Gaussians, four isotropic clusters at \( (\pm 2, 0) \) and \( (0, \pm 2) \) with \( \sigma = 0.2 \);
    \textbf{(b)} 2 Moons, two interleaved half-moon arcs with additive Gaussian noise;
    \textbf{(c)} Swiss Roll, a planar spiral manifold.
    \textit{Top row:} snapshots of the learned potential \( W(t, \mathbf{x}) \) at \( t = 0, 0.5, 1.0 \), showing the emergence of structured basins aligned with the target geometry as transport progresses.
    \textit{Middle row:} particle positions at each time step under the learned backward-time controlled diffusion (Algorithm~\ref{alg:HJB_generation}), demonstrating successful aggregation onto \( p_{\rm data} \) at terminal time \( t = 1 \).
    \textit{Bottom row:} convergence of \( \mathcal{L}_{\rm total} \) over training epochs; monotonic decay across all three datasets confirms robust Feynman--Kac trajectory supervision without explicit score estimation or backward-time SDE simulation.}
    \label{fig:example}
\end{figure}

We present results demonstrating the capabilities of our method across three axes: (i) training performance for the HJB value function Feynman--Kac trajectory supervision, (ii) generative modeling performance, and (iii) the influence of the learned cost geometry \( \nu(x) \) on shaping transport paths. These experiments collectively validate the theoretical underpinnings of our framework and illustrate its practical effectiveness in learning minimal-cost stochastic transport scalar potential function from data.

% \vspace{-5pt}
\paragraph{Training the HJB value function via Feynman--Kac trajectory supervision.}
We first tested our method on three standard 2D benchmark datasets: 4 Gaussians, 2 Moons, and Swiss Roll (Fig.~\ref{fig:example}). The reference distribution is an isotropic Gaussian \( p_{\rm ref} \sim \mathcal{N}(0, I) \) and the spatial cost field is initialized uniformly, \( \nu(\mathbf{x}) = 1 \). The value function \( W(t, \mathbf{x}) \) is parameterized as a 10-layer MLP with 64 hidden units per layer, GeLU activations, and sinusoidal positional time embeddings. Forward trajectories are generated by an Ornstein--Uhlenbeck process (\( D = 0.05 \), \( \beta = 0.1 \)) discretized over \( T = 128 \) steps, and the model is trained using \( \mathcal{L}_{\rm total} = \mathcal{L}_{\rm FK} + \mathcal{L}_{\rm dual} \) (Eqn.~\ref{eq:total_loss}, Algorithm~\ref{alg:HJB_training}) for \( n = 2000 \) epochs on \( N = 4096 \) samples per dataset (Adam, \( \eta = 10^{-3} \), JAX, single NVIDIA GPU, float32).

As training progresses, the value function develops structured basins that reflect the geometry of \( p_{\rm data} \). For the 4 Gaussians dataset, four distinct potential wells emerge at \( t = 1 \), each centered on a cluster. For 2 Moons, the potential organizes into two arc-shaped valleys tracing the half-moon arcs. For Swiss Roll, it develops a spiraling basin aligned with the manifold. This structure emerges purely from forward-time trajectory supervision --- no backward SDE simulation or score estimation is performed. The learned drift \( \mathbf{u}^* = \nabla_{\mathbf{x}} W \) then guides backward-time sampling (Algorithm~\ref{alg:HJB_generation}), steering particles from \( p_{\rm ref} \) to \( p_{\rm data} \) along minimal-cost paths. Monotonic decay of \( \mathcal{L}_{\rm total} \) across all three datasets (Fig.~\ref{fig:example}, bottom row) confirms the stability of the Feynman--Kac estimator across qualitatively different target geometries.

% \vspace{-10pt}
\paragraph{Validation of Action-Minimizing Ground States.}
We evaluate the capacity of our formulation to discover transport plans mapping \( p_{\rm ref} \sim \mathcal{N}(0, I) \) to lower-entropy empirical targets with qualitatively different topological structure. The evolution of \( W \) in Fig.~\ref{fig:example} reveals time-dependent basins that directly reflect the geometry of each \( p_{\rm data} \): four symmetric potential wells for 4 Gaussians, two arc-shaped troughs for 2 Moons, and a spiraling channel for Swiss Roll. These basins form progressively from \( t = 0 \) to \( t = 1 \), acting as dynamically evolving attractors that guide transported particles. Particles initialized from \( p_{\rm ref} \) successfully aggregate onto the target manifold at terminal time \( t = 1 \), following paths that minimize the combined spatial and control cost without any topology-specific inductive bias or architectural change. The topological diversity of the benchmarks --- disconnected clusters, non-convex arcs, and a multiply-wound spiral --- demonstrates that the HJB potential adapts its geometry to each target through the learned value function alone, validating the coherence of our forward--backward duality as a general-purpose transport mechanism.

\begin{figure}[htpb]
    \centering
    \includegraphics[width=1.0\linewidth]{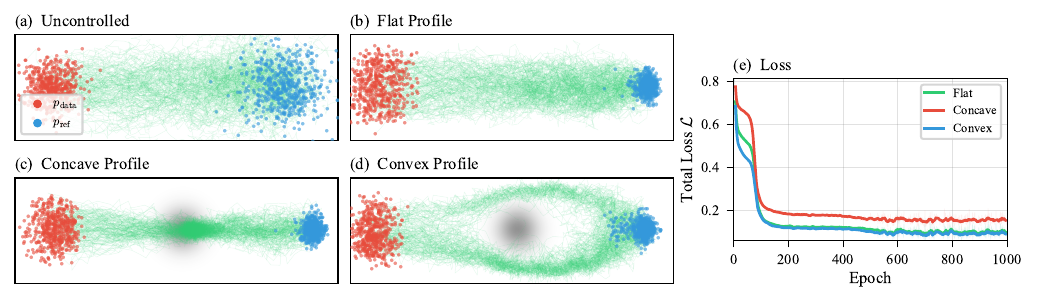}
    \caption{\small
    \textbf{Geometric control of stochastic transport via spatial cost field \(\nu(x)\).} We solve the same stochastic optimal control problem as in Fig.~\ref{fig:example} (Eqn.~\ref{eq:generative_optimal_ctrl_formulation}), but with an inhomogeneous spatial cost field \(\nu(x)\) that shapes the geometry of optimal transport paths (cf. Fermat's principle). The source is a Gaussian \( p_0 = \mathcal{N}((-1,0),\, 0.01\,I) \) and the target is \( p_1 = \mathcal{N}((1,0),\, 0.01\,I) \), with a Gaussian \(\nu(x)\) profile (mean at origin, \(\sigma_\nu = 0.1\)) placed at the midpoint between them. The value function \(W(t,\mathbf{x})\) is parameterized as a 10-layer MLP (64 hidden units, sinusoidal time embeddings) and trained via Feynman--Kac trajectory supervision (Algorithm~\ref{alg:HJB_training}, \( \mathcal{L}_{\mathrm{total}} = \mathcal{L}_{\mathrm{FK}} + \mathcal{L}_{\mathrm{dual}} \), Eqn.~\ref{eq:total_loss}). Forward trajectories use an Ornstein--Uhlenbeck process (\( D = 0.05 \), \( \beta = 0.1 \)) with \( T = 100 \) steps; training runs for \( n = 1000 \) epochs on \( N = 512 \) samples (Adam, \( \eta = 10^{-3} \), JAX). Trajectories (green) connect \( p_0 \) (red) to \( p_1 \) (blue). \textbf{(a)}~Uncontrolled forward diffusion (no learned control), producing unstructured spreading. \textbf{(b)}~Controlled transport with flat \(\nu(x) = \mathrm{const}\), recovering straight-line baseline geometry. \textbf{(c)}~Concave cost profile: low \(\nu\) at the midpoint creates a potential well, concentrating optimal paths inward through the center. \textbf{(d)}~Convex cost profile: high \(\nu\) at the midpoint acts as an energetic barrier, deflecting optimal paths outward around the obstacle. \textbf{(e)}~Convergence of \(\mathcal{L}_{\mathrm{total}}\) for the three controlled profiles, confirming robust training under different spatial cost fields.}
    \label{fig:refract}
\end{figure}

% \vspace{-5pt}
\paragraph{Geometry of Refraction and Fermat's Principle for Stochastics.}
A distinguishing feature of our framework is the spatial cost field \( \nu(\mathbf{x}) \), which enters the HJB equation (Eqn.~\ref{eq:forward_hjb}) as a running cost and directly shapes the geometry of optimal transport paths. To isolate this effect cleanly, we consider a controlled 2D experiment: a Gaussian source \( p_0 = \mathcal{N}((-1,0),\, 0.01\,I) \) transported to a Gaussian target \( p_1 = \mathcal{N}((1,0),\, 0.01\,I) \), with a localized Gaussian \(\nu(\mathbf{x})\) profile (\( \sigma_\nu = 0.1 \)) placed at the midpoint of the transport path (Fig.~\ref{fig:refract}).

The physical analogy is direct and precise. In classical optics, Fermat's Principle of least time dictates that light follows the path of minimum optical length in a medium with spatially varying refractive index \( n(\mathbf{x}) \), corresponding to geodesics of the Riemannian metric \( n(\mathbf{x})^2 \langle \cdot, \cdot \rangle \). Here, \( \nu(\mathbf{x}) \) plays an exactly analogous role: regions of high \( \nu \) increase the local running cost of transiting that region, raising the value function \( W \) and deflecting optimal trajectories away --- analogous to a denser optical medium refracting light outward. Conversely, regions of low \( \nu \) create a potential well in \( W \) that attracts and focuses trajectories --- analogous to a converging lens. The HJB value function thus encodes the geometry of these least-cost paths precisely as a physical path-space free energy modulated by the local metric defined by \( \nu \).

We demonstrate this with three controlled configurations (Fig.~\ref{fig:refract}, panels (b--d)). A \emph{flat} profile (\( \nu = \mathrm{const} \)) recovers straight-line baseline geometry between source and target. A \emph{convex cost lens}, a Gaussian peak in \( \nu \) at the midpoint --- acts as an energetic barrier: optimal paths bend outward around the obstacle, reproducing diverging optical behavior. A \emph{concave cost lens}, a Gaussian trough in \( \nu \) at the midpoint, acts as a potential well: paths are drawn inward and concentrate through the center, reproducing converging optical behavior. The uncontrolled forward diffusion (leftmost panel) exhibits unstructured spreading with no preferred direction, confirming that all observed deflections arise from the learned HJB potential shaped by \( \nu(\mathbf{x}) \) alone.

Crucially, all four configurations use identical neural architectures, training algorithms, and SDE dynamics, only \( \nu(\mathbf{x}) \) changes. This demonstrates that \( \nu(\mathbf{x}) \) acts as a powerful and interpretable geometric prior on the transport field, enabling deterministic confinement, deflection, or focusing of stochastic trajectories through spatial cost design alone. From a practical standpoint, this opens a route to constrained or guided generation: by shaping \( \nu(\mathbf{x}) \) to encode physical, geometric, or domain constraints, one can steer learned generative dynamics into prescribed regions of state space without modifying the underlying model.

\section{Validation of PDE Residuals and Path Variance}
A common structural concern in applying Cole-Hopf linearization techniques to empirical distributions is verifying whether the parameterized potential strictly obeys the underlying mathematical theory globally. The primary measurable observables of our theoretical bounds are the magnitude of the empirical PDE residual of the forward HJB equation, $|\frac{\partial W}{\partial s} - D \Delta W - \frac{1}{2\gamma} \| \nabla W \|^2 + \nu(x)|$, and the true empirical variance of the Feynman-Kac path-integral estimator evaluated as the physical state-space coordinates scale. While our formulation guarantees exact analytical equivalence in the theoretical limit of infinite unbiased sampling, scrutinizing the explicit statistical growth of the systematic bias and the variance stability over high-dimensional empirical potentials remains a critical task for future analytical evaluations. Measuring these physical deviation limits natively against standard macroscopic Schrödinger Bridge models will directly quantify whether synthesized vector fields perfectly guarantee trajectory optimality limits.

% % \vspace{-10pt}
\section{Conclusions}
We have presented a principled theoretical framework for controlling non-equilibrium multibody physics systems through a matched pair of forward- and backward-time Hamilton–Jacobi–Bellman (HJB) equations. This formulation mathematically extends existing optimal control perspectives on diffusion processes by deriving a trajectory-based HJB formulation that integrates an explicit spatially-dependent transport geometry. Inside this formulation, the physical drift is analytically obtained as the gradient of a scalar value function, while the time-reversed evolution yields a forward HJB equation exactly evaluated by a generalized Feynman–Kac representation. This proof enables the free energy—and hence the ensemble dynamics—to be reconstructed directly from forward stochastic trajectories without performing unreliable backward simulation approximations. score estimation or backward-time simulation.

This perspective differs from most existing generative transport approaches as it begins from a stochastic optimal control problem with an explicit pathwise cost functional. The generative dynamics arise from the value function solving the associated HJB equation, providing a direct variational characterization of the transport process. A further distinguishing feature is the role of the spatial cost field, which defines a geometry on state space that shapes admissible transport trajectories. In this sense, the learned value function encodes the least-cost paths of a controlled stochastic system rather than merely a density transformation. The formulation therefore emphasizes trajectory-level structure and control effort, rather than purely distributional matching.

More broadly, the forward–backward Hamilton–Jacobi correspondence highlights structural links between stochastic optimal control, Schrodinger bridge theory, and generative transport. These connections suggest that generative dynamics may be understood as controlled diffusions governed by variational principles on path space. Extending this perspective to higher-dimensional systems, interacting particles, and learned transport geometries may provide new tools for modeling and controlling stochastic processes in complex physical and biological systems. Since the cost landscape  defines the geometry of transport, our approach offers a way to confine generative dynamics within prescribed regions of state space by fine-tuning pretrained models through spatially structured cost fields that are guided by physics and control priors.

\paragraph{Code availability.}
The source code for all experiments presented in this work is publicly available at \url{https://github.com/sumit-sinha-seas/HJB_matching.git}.

\clearpage
\bibliographystyle{apsrev4-2}
\bibliography{references}

\clearpage
{\Large \textbf{Appendix}}
\appendix

\section{Proof of Lemma~\ref{lemma:generative_opt_ctrl}} \label{app:proof:generative_opt_ctrl}
Let \( p(t, \mathbf{x}) \) denote the time-marginal density of the process \( \mathbf{x}_t \) governed by the controlled SDE
\[
d\mathbf{x}_t = \mathbf{u}(t, \mathbf{x}_t)\,dt + \sqrt{2D} \, d\mathbf{B}_t, \quad \mathbf{x}_0 \sim p_{\mathrm{ref}}, \quad \mathbf{x}_1 \sim p_{\mathrm{data}}.
\]
The evolution of \( p(t, \mathbf{x}) \) satisfies the Fokker--Planck equation
\[
\frac{\partial p}{\partial t} = - \nabla \cdot (p \mathbf{u}) + D \Delta p, \quad p(0, \mathbf{x}) = p_{\mathrm{ref}}(\mathbf{x}), \quad p(1, \mathbf{x}) = p_{\mathrm{data}}(\mathbf{x}).
\]

The expected total cost is
\[
\mathcal{J}[p, \mathbf{u}] = \int_0^1 \int_{\mathbb{R}^d} p(t, \mathbf{x}) \left[ \nu(\mathbf{x}) + \frac{\gamma}{2} \| \mathbf{u}(t, \mathbf{x}) \|^2 \right] d\mathbf{x} \, dt.
\]

To enforce the Fokker--Planck constraint, we introduce a scalar Lagrange multiplier \( U(t, \mathbf{x}) \) and define the Lagrangian
\[
\mathcal{L}[p, \mathbf{u}, U] = \mathcal{J}[p, \mathbf{u}] - \int_0^1 \int_{\mathbb{R}^d} U(t, \mathbf{x}) \left( \frac{\partial p}{\partial t} + \nabla \cdot (p \mathbf{u}) - D \Delta p \right) d\mathbf{x} \, dt.
\]

Assuming sufficient regularity and rapid decay at infinity, we integrate by parts to move derivatives off \( p \)
\[
\begin{aligned}
\mathcal{L} =\; &\int_0^1 \int_{\mathbb{R}^d} p(t, \mathbf{x}) \left[ \nu(\mathbf{x}) + \frac{\gamma}{2} \| \mathbf{u}(t, \mathbf{x}) \|^2 + \frac{\partial U}{\partial t} + \nabla U \cdot \mathbf{u} + D \Delta U \right] d\mathbf{x} \, dt \\
&\quad + \int_{\mathbb{R}^d} U(0, \mathbf{x}) \, p_{\mathrm{ref}}(\mathbf{x}) \, d\mathbf{x} - \int_{\mathbb{R}^d} U(1, \mathbf{x}) \, p_{\mathrm{data}}(\mathbf{x}) \, d\mathbf{x}.
\end{aligned}
\]

We now minimize \( \mathcal{L} \) over \( \mathbf{u} \). For fixed \( p, U \), this is a pointwise minimization
\[
\min_{\mathbf{u}} \left\{ \frac{\gamma}{2} \| \mathbf{u} \|^2 + \nabla U \cdot \mathbf{u} \right\},
\]
with unique minimizer
\[
\mathbf{u}^*(t, \mathbf{x}) = - \frac{1}{\gamma} \nabla U(t, \mathbf{x}).
\]

Substituting \( \mathbf{u}^* \) into \( \mathcal{L} \) gives
\[
\mathcal{L}[p, U] = \int_0^1 \int p \left[ \nu + \frac{\partial U}{\partial t} + D \Delta U - \frac{1}{2\gamma} \| \nabla U \|^2 \right] d\mathbf{x} \, dt + \text{boundary terms}.
\]

To eliminate the interior integral, the integrand must vanish, yielding the Hamilton--Jacobi--Bellman (HJB) equation
\[
\frac{\partial U}{\partial t} + D \Delta U - \frac{1}{2\gamma} \| \nabla U \|^2 + \nu = 0.
\]

Maximizing over boundary values of \( U \) gives the dual variational problem
\begin{align*}
\max_{U(0), U(1)} &\left\{ \int U(0, \mathbf{x}) \, p_{\mathrm{ref}}(\mathbf{x}) d\mathbf{x} - \int U(1, \mathbf{x}) \, p_{\mathrm{data}}(\mathbf{x}) d\mathbf{x} \right\}
\\ &\text{s.t.} \quad \frac{\partial U}{\partial t} + D \Delta U - \frac{1}{2\gamma} \| \nabla U \|^2 + \nu = 0.
\end{align*}
This completes the proof.

\section{Proof of Theorem~\ref{thm:fwd-bwd-hjb}} \label{app:fwd-bwd-hjb}
Let \( U(t, \mathbf{x}) \) be the solution to the backward-time Hamilton--Jacobi--Bellman (HJB) equation from Lemma~\ref{lemma:generative_opt_ctrl}, and define the forward-time potential \( W(s, \mathbf{x}) := -U(1 - s, \mathbf{x}) \). We now prove each part in turn.

\vspace{5pt}
\noindent\textbf{(1) Forward HJB equation.}

Differentiating \( W(s, \mathbf{x}) = -U(1 - s, \mathbf{x}) \) with respect to \( s \), we obtain
\[
\frac{\partial W}{\partial s}(s, \mathbf{x}) = \frac{\partial U}{\partial t}(1 - s, \mathbf{x}).
\]
Using the HJB satisfied by \( U \)
\[
\frac{\partial U}{\partial t} + D \Delta U - \frac{1}{2\gamma} \| \nabla U \|^2 + \nu(\mathbf{x}) = 0,
\]
we evaluate at \( t = 1 - s \) and substitute in the previous equation to get
\[
\frac{\partial W}{\partial s}(s, \mathbf{x}) = - D \Delta U(1 - s, \mathbf{x}) + \frac{1}{2\gamma} \| \nabla U(1 - s, \mathbf{x}) \|^2 - \nu(\mathbf{x}).
\]
Note that \( \nabla W(s, \mathbf{x}) = -\nabla U(1 - s, \mathbf{x}) \), so \( \| \nabla W \|^2 = \| \nabla U \|^2 \) and \( \Delta W = - \Delta U \). Therefore, we get
\[
\frac{\partial W}{\partial s} - D \Delta W - \frac{1}{2\gamma} \| \nabla W \|^2 + \nu(\mathbf{x}) = 0,
\]
which proves that \( W \) satisfies the forward HJB~\eqref{eq:forward_hjb}.

\vspace{10pt}
\noindent\textbf{(2) The forward controlled process transports \( p_{\mathrm{data}} \to p_{\mathrm{ref}} \).}

Consider the controlled forward SDE
\[
d\mathbf{y}_s = \mathbf{v}^*(s, \mathbf{y}_s) \, ds + \sqrt{2D} \, d\mathbf{B}_s,
\quad \text{with} \quad
\mathbf{v}^*(s, \mathbf{x}) = -\frac{1}{\gamma} \nabla W(s, \mathbf{x}) + 2D \nabla \log q(s, \mathbf{x}),
\]
where \( q(s, \mathbf{x}) \) is the time-marginal density of \( \mathbf{y}_s \), initialized at \( \mathbf{y}_0 \sim p_{\mathrm{data}} \). The Fokker--Planck equation for \( q \) is given by
\[
\frac{\partial q}{\partial s} = -\nabla \cdot (q \mathbf{v}^*) + D \Delta q.
\]
First, we note that
\[
q \mathbf{v}^* = -\frac{1}{\gamma} q \nabla W + 2D q \nabla \log q = -\frac{1}{\gamma} q \nabla W + 2D \nabla q.
\]
Therefore, substituting for \( q \mathbf{v}^* \) in the Fokker-Planck equation for $q$, we get
\[
\frac{\partial q}{\partial s} 
= -\nabla \cdot \left( -\frac{1}{\gamma} q \nabla W + 2D \nabla q \right) + D \Delta q
= \frac{1}{\gamma} \nabla \cdot (q \nabla W) - D \Delta q.
\]
Now define the backward-time marginal density \( p(t, \mathbf{x}) := q(1 - t, \mathbf{x}) \). Then, we obtain
\[
\frac{\partial p}{\partial t} = - \left. \frac{\partial q}{\partial s} \right|_{s = 1 - t}
= -\left( \frac{1}{\gamma} \nabla \cdot (q \nabla W) - D \Delta q \right)_{s = 1 - t}
= D \Delta p - \frac{1}{\gamma} \nabla \cdot \left( p \nabla W(1 - t, \mathbf{x}) \right).
\]
Futhermore, using \( W(1 - t, \mathbf{x}) = -U(t, \mathbf{x}) \), we find
\[
\frac{\partial p}{\partial t} = D \Delta p + \frac{1}{\gamma} \nabla \cdot (p \nabla U),
\]
which is the Fokker--Planck equation corresponding to the backward-time controlled SDE
\[
d\mathbf{x}_t = -\frac{1}{\gamma} \nabla U(t, \mathbf{x}_t) \, dt + \sqrt{2D} \, d\mathbf{B}_t.
\]
We know from Lemma~\ref{lemma:generative_opt_ctrl} that the above process satisfies the boundary conditions $\mathbf{x}_0 \sim p_{\rm ref}$ and $\mathbf{x}_1 \sim p_{\rm data}$, i.e., $p(0, \mathbf{x}) = p_{\rm ref}(\mathbf{x})$ and $p(1, \mathbf{x}) = p_{\rm data}(\mathbf{x})$. We now have
\[
p(1, \mathbf{x}) = q(0, \mathbf{x}) = p_{\mathrm{data}}(\mathbf{x}), \quad p(0, \mathbf{x}) = q(1, \mathbf{x}) = p_{\mathrm{ref}}(\mathbf{x}).
\]
Hence, the process \( \mathbf{y}_s \) evolves from \( p_{\mathrm{data}} \to p_{\mathrm{ref}} \), and is the time-reversal of the optimal process governed by \( U \).

\vspace{10pt}
\noindent\textbf{(3) Optimality of the forward control \( \mathbf{v}^* \).}

We now show that \( \mathbf{v}^* \) minimizes the objective
\[
\mathcal{J}[\mathbf{v}] := \mathbb{E}_{\mathbb{Q}_{\mathbf{v}}} \left[ \int_0^1 \nu(\mathbf{y}_s) \, ds + \frac{\gamma}{2} \int_0^1 \| \mathbf{v}_s - \mathbf{s}_s \|^2 \, ds \right],
\]
subject to \( d\mathbf{y}_s = \mathbf{v}_s \, ds + \sqrt{2D} \, d\mathbf{B}_s \), with \( \mathbf{y}_0 \sim p_{\mathrm{data}} \) and \( \mathbf{y}_1 \sim p_{\mathrm{ref}} \), where \( \mathbf{s}_s := 2D \nabla \log q(s, \mathbf{x}) \). We first note that $\| \mathbf{v}_s - \mathbf{s}_s \|^2 
= \| \mathbf{v}_s \|^2 - 2 \mathbf{v}_s \cdot \mathbf{s}_s + \| \mathbf{s}_s \|^2$. Substituting in the expression for $\mathcal{J}[\mathbf{v}]$, we get
\[
\mathcal{J}[\mathbf{v}] 
= \mathbb{E}_{\mathbb{Q}_{\mathbf{v}}} \left[ \int_0^1 \nu(\mathbf{y}_s) + \frac{\gamma}{2} \| \mathbf{v}_s \|^2 - \gamma \mathbf{v}_s \cdot \mathbf{s}_s + \frac{\gamma}{2} \| \mathbf{s}_s \|^2 \, ds \right].
\]
We now express this as an integral with respect to the marginal probability density \( q(s, \mathbf{x}) \) as
\[
\mathcal{J}[\mathbf{v}] = \int_0^1 \int_{\mathbb{R}^d} q(s, \mathbf{x}) \left( \nu(\mathbf{x}) + \frac{\gamma}{2} \| \mathbf{v}(s, \mathbf{x}) \|^2 - \gamma \mathbf{v}(s, \mathbf{x}) \cdot \mathbf{s}(s, \mathbf{x}) + \frac{\gamma}{2} \| \mathbf{s}(s, \mathbf{x}) \|^2 \right) d\mathbf{x} \, ds.
\]
We note that the above objective function is subject to the forward SDE constraint. We construct the corresponding Lagrangian using the associated Fokker-Planck equation for~$q(s, \mathbf{x})$, given by $\frac{\partial q}{\partial s} = -\nabla \cdot (q \mathbf{v}) + D \Delta q$,
with Lagrange multiplier \( W(s, \mathbf{x}) \) as follows
\begin{align*}
\mathcal{L}[q, \mathbf{v}, W] 
= \mathcal{J}[\mathbf{v}] 
- \int_0^1 \int_{\mathbb{R}^d} W \left( \frac{\partial q}{\partial s} + \nabla \cdot (q \mathbf{v}) - D \Delta q \right) d\mathbf{x} \, ds.
\end{align*}
Integrating by parts, assuming smooth decay at infinity, we obtain
\begin{align*}
\mathcal{L} = &\int W(0, \mathbf{x}) q(0, \mathbf{x}) d\mathbf{x} - \int W(1, \mathbf{x}) q(1, \mathbf{x}) d\mathbf{x} \\
&+ \int_0^1 \int q \left( \nu + \frac{\gamma}{2} \| \mathbf{v} \|^2 - \gamma \mathbf{v} \cdot \mathbf{s} + \frac{\gamma}{2} \| \mathbf{s} \|^2 + \frac{\partial W}{\partial s} + \nabla W \cdot \mathbf{v} + D \Delta W \right) d\mathbf{x} ds.
\end{align*}
Minimizing \( \mathcal{L} \) pointwise with respect to \( \mathbf{v} \) gives $\mathbf{v}^* = -\frac{1}{\gamma} \nabla W + \mathbf{s}$, and substituting into the integrand, we retrieve the forward HJB equation
\[
\frac{\partial W}{\partial s} - D \Delta W - \frac{1}{2\gamma} \| \nabla W \|^2 + \nu = 0.
\]
This confirms that \( \mathbf{v}^* = -\frac{1}{\gamma} \nabla W + 2D \nabla \log q \) is the minimizer of the stated cost functional, completing the proof.

\section{Comments on Feynman-Kac estimation of the forward potential} \label{app:forward-potential-clarifications}
This appendix expands on the theoretical structure underlying the Feynman--Kac representation of the forward potential \( W \), introduced in Section~\ref{sec:fwd_rev_hjb_matching}. While the main derivation establishes the connection between stochastic optimal control and the forward HJB equation, several subtleties warrant further discussion.

We begin by addressing the efficiency of path sampling using drifted reference processes and the necessary Girsanov reweighting. We then show that the control objective encodes not only expected path cost but also variance minimization, revealing a natural risk-sensitive interpretation. Finally, we present a regularized approach for learning the cost function \( \nu(x) \) from trajectory densities, grounded in the Euler--Lagrange structure of the objective functional.

\subsection{Efficient sampling employing Girsanov correction}
In practice, sampling trajectories from pure Brownian motion (as assumed in the original Feynman–Kac formulation) can be inefficient when the data distribution \( p_{\mathrm{data}} \) is far from the reference \( p_{\mathrm{ref}} \). To address this, we simulate forward trajectories using a drifted reference process, typically a Langevin dynamics of the form
\begin{align} \label{eq:langevin_app}
    d\mathbf{x}_s = -\nabla V(\mathbf{x}_s) \, ds + \sqrt{2D} \, d\mathbf{B}_s, \quad \mathbf{x}_0 \sim p_{\mathrm{data}}.
\end{align}
This dynamics induces a path measure \( \mathbb{P}_V \) different from the Brownian baseline \( \mathbb{P}_0 \). Girsanov's theorem provides a principled way to reweight trajectories sampled under \( \mathbb{P}_V \) to obtain unbiased estimates under \( \mathbb{P}_0 \). Specifically, the Radon--Nikodym derivative between these measures is
\[
    \frac{d\mathbb{P}_0}{d\mathbb{P}_V} = \exp\left( \frac{1}{2D} \int_0^t \nabla V(\mathbf{x}_s) \cdot d\mathbf{x}_s + \frac{1}{4D} \int_0^t \| \nabla V(\mathbf{x}_s) \|^2 \, ds \right).
\]
Substituting into the Feynman–Kac representation yields an unbiased estimator for \( Z(t, \mathbf{x}) \) under Langevin sampling:
\begin{align*}
\small
\begin{aligned}
    Z(t, \mathbf{x}) = \mathbb{E}_{\mathbb{P}_V} \left[ \left. Z(0, \mathbf{x}_0) \exp\left( - \beta \int_0^t \nu(\mathbf{x}_s) \, ds + \frac{1}{2D} \int_0^t \nabla V(\mathbf{x}_s) \cdot d\mathbf{x}_s + \frac{1}{4D} \int_0^t \| \nabla V(\mathbf{x}_s) \|^2 \, ds \right) \,\right|\, \mathbf{x}_t = \mathbf{x} \right].
\end{aligned}
\end{align*}
If the drift due to the potential~$V$ is not corrected for in the backward generative process, this correction allows the use of more efficient reference dynamics for forward path sampling while still preserving theoretical correctness.

\subsection{Risk-Sensitive Interpretation and Variance Control}
The generative potential \( U(t, \mathbf{x}) \) defined in Lemma~\ref{lemma:generative_opt_ctrl} satisfies a backward Hamilton–Jacobi–Bellman (HJB) equation, and admits a Feynman–Kac representation that evaluates expectations over path space starting from time \( t \). In particular, letting \( \beta = \nicefrac{1}{2\gamma D} \), we can express \( U \) as
\[
U(t, \mathbf{x}) = - \frac{1}{\beta} \log \mathbb{E}_{\mathbb{P}_0} \left[ \left. \exp\left( - \beta \left( \int_t^1 \nu(\mathbf{x}_s) \, ds + U(1, \mathbf{x}_1) \right) \right) \right| \mathbf{x}_t = \mathbf{x} \right],
\]
where \( \mathbf{x}_s \) follows Brownian motion. Expanding this expression via Laplace's method (or a cumulant expansion), we obtain
\[
U(t, \mathbf{x}) = \mathbb{E}[C \mid \mathbf{x}_t = \mathbf{x}] - \frac{\beta}{2} \operatorname{Var}[C \mid \mathbf{x}_t = \mathbf{x}] + \mathcal{O}(\beta^2),
\]
where
\[
C := \int_t^1 \nu(\mathbf{x}_s) ds + U(1, \mathbf{x}_1)
\]
denotes the total trajectory cost-to-go. This reveals that the control objective implicitly incorporates both the expected cost and its variance. The parameter \( \gamma \) (via \( \beta = 1 / (2D \gamma) \)) controls this tradeoff: large \( \gamma \) corresponds to risk-neutral (variance-tolerant) control, while small \( \gamma \) induces risk-averse behavior, concentrating trajectories on lower-cost, more deterministic paths. This form of variance control arises naturally from the stochastic control formulation and obviates the need for explicit regularization.

\section{Scalability to high-dimensional systems.}
\begin{figure}[htpb]
    \centering
    \includegraphics[width=0.98\linewidth]{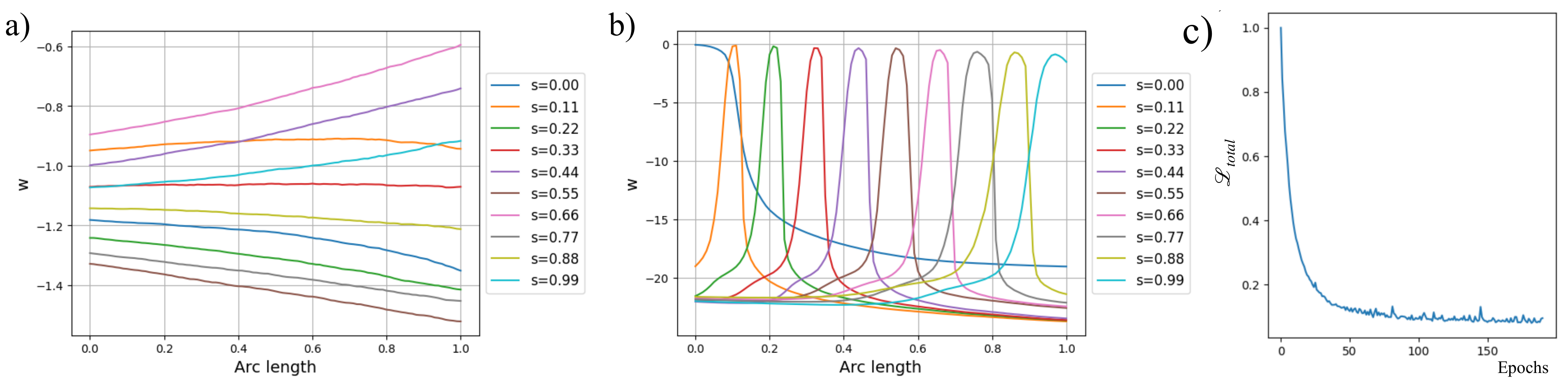}
    \caption{\small
    \textbf{HJB value function learning in high dimensions: transport from Gaussian noise to MNIST.}
    We solve the same stochastic optimal control problem (Eqn.~\ref{eq:generative_optimal_ctrl_formulation}) as in the 2D setting, now in the high-dimensional pixel space of MNIST (\(28 \times 28 = 784\) dimensions). The reference \( p_{\rm ref} = \mathcal{N}(0, I) \) is transported to \( p_{\rm data} \) (MNIST digit images) with a fixed uniform cost field \( \nu(\mathbf{x}) = 1 \).
    The value function \( W_\theta(t, \mathbf{x}) \) is parameterized as a convolutional U-Net (encoder channels \([32, 64, 128, 256]\), GroupNorm, skip connections, Gaussian Fourier time embeddings with \(\mathrm{embed\_dim} = 256\)) and trained via Feynman--Kac trajectory supervision using the total loss \( \mathcal{L}_{\rm total} = \mathcal{L}_{\rm FK} + \mathcal{L}_{\rm dual} + \mathcal{L}_{\rm spatial} \) (Eqn.~\ref{eq:total_loss}, Algorithm~\ref{alg:HJB_training}).
    Forward trajectories are sampled from the closed-form conditional of an Ornstein--Uhlenbeck process (\( D = 0.01 \), \( \theta = 5.0 \), \( \beta = 0.1 \), \( T = 100 \) steps), enabling memory-efficient training without rolling out full forward trajectories.
    Training runs for \( n = 200 \) epochs on the full MNIST training set (\(\text{batch size} = 1024\)) using the Adam optimizer (\( \eta = 10^{-4} \)), implemented in PyTorch.
    We visualize the scalar potential \( W(t, \mathbf{x}(s)) \) evaluated along held-out test trajectories \( \{ \mathbf{x}(s) \} \) (unseen during training) at successive time steps.
    \textbf{(a)} At initialization, the potential is unstructured and shows no coherent alignment with the transport direction.
    \textbf{(b)} After 200 epochs, a clear propagating bump emerges: each trajectory exhibits a moving pulse in \( W \) aligned with the generative path, consistent with the learned cost-to-go structure observed in the 2D setting (Fig.~\ref{fig:example}).
    \textbf{(c)} Convergence of \( \mathcal{L}_{\rm total} \) over training epochs. The emergence of coherent potential structure on test trajectories confirms that the Feynman--Kac framework generalizes beyond the training ensemble to high-dimensional image data.}
    \label{fig:MNIST}
\end{figure}

\noindent To assess whether the learned potential exhibits similar structural coherence in high-dimensional empirical ensembles, we apply our framework to transport Gaussian noise \( p_{\rm ref} = \mathcal{N}(0, I) \) to the MNIST handwritten digit dataset~\cite{lecun2010mnist} (\( 28 \times 28 = 784 \) dimensions, Figure~\ref{fig:MNIST}). A fixed uniform cost field \( \nu(\mathbf{x}) = 1 \) in~\eqref{eq:generative_optimal_ctrl_formulation} is used throughout. Rather than rolling out full forward trajectories, we exploit the closed-form conditional of the Ornstein--Uhlenbeck process (\( D = 0.01 \), \( \theta = 5.0 \), \( \beta = 0.1 \)) to sample perturbed states \( \mathbf{x}_t \) directly at any time \( t \), enabling memory-efficient training without storing complete trajectory arrays. The value function is parameterized as a convolutional U-Net with encoder channels \([32, 64, 128, 256]\), GroupNorm, dense skip connections, and Gaussian Fourier time embeddings (\(\mathrm{embed\_dim} = 256\)). The model is trained using an augmented loss \( \mathcal{L}_{\rm total} = \mathcal{L}_{\rm FK} + \mathcal{L}_{\rm dual} + \mathcal{L}_{\rm spatial} \) for \( n = 200 \) epochs on the full MNIST training set (batch size 1024, Adam, \( \eta = 10^{-4} \), PyTorch).

We visualize \( W(t, \mathbf{x}(s)) \) evaluated along held-out test trajectories \( \{\mathbf{x}(s)\} \) unseen during training. At initialization, the potential is unstructured and shows no coherent alignment with the transport direction. After 200 epochs, a clear propagating bump emerges: each trajectory exhibits a moving pulse in \( W \) aligned with the generative path, consistent with the learned cost-to-go structure. This behavior mirrors the 2D case (Fig.~\ref{fig:example}), where the potential evolves as a coherent pulse that steers mass toward the data manifold. The appearance of this structure on test trajectories --- not seen during training --- demonstrates that the Feynman--Kac framework learns a globally consistent value function, not merely a local interpolation over training samples. The loss convergence (Fig.~\ref{fig:MNIST}, bottom) confirms stable optimization in the high-dimensional image setting, and the thermodynamic interpretability of the HJB potential is preserved across dimensionalities.

\end{document}